\documentclass{article}
\usepackage{preprint/authblk}
\usepackage{graphicx}
\usepackage{cite}
\begin{document}

\title{Lightweight Record-and-Replay for Intermittent Tests Failures}
\author[1]{Omar S Navarro Leija}
\author[2]{Alan Jeffrey}
\affil[1]{Mozilla Research Intern, University Of Pennsylvania}
\affil[2]{Mozilla Corporation}
\date{Summer 2019}
\maketitle

\begin{abstract}
In this paper we present lightweight record-and-replay (RR). In contrast to traditional ``fully deterministic'' RR solutions, lightweight RR focuses on handling nondeterminism arising from thread communication for programs with concurrent, message-passing architectures. By decreasing nondeterminism in programs, lightweight RR decreases the number of intermittent failures in program's test suites. We evaluated the effectiveness of lightweight RR on Servo, a highly concurrent web browser. Our evaluation shows lightweight RR is effective at greatly reducing intermittent failures for some tests, but not others. Lightweight RR performance overhead remains a work in progress, but log sizes are quite small. We believe with further work lightweight RR could prove useful for lowering nondeterminism in programs at a negligible performance overhead.
\end{abstract}

\section{Background}

\subsection{Concurrent Message Passing Channels}
Slowdowns of single-core year-to-year performance gains in computing systems have given rise to the need
for highly concurrent systems that take advantage of multiple cores, now available
in all machines.
Concurrent systems are normally implemented using a thread plus shared-memory,
using locks to enforce exclusive writes to memory. This type of systems make
it difficult to reason about useful properties of concurrent programs:
data race freedom and deadlock avoidance.
Furthermore shared data and the locks that protect it are often logically decoupled. That is,
grabbing a lock before accessing data is not enforced by the programming language or runtime.
Instead, in the worst case, it is merely a programming convention buried in the documentation,
or at best, enforced through the code API and encapsulation.

An alternative concurrency model popularized by languages like Go, Erlang, and Rust has
arisen: message passing through channels. With message passing, isolated
threads or processes communicate via channels, sending data as messages to each other.
This avoids the need for the programmer to worry about data races, memory consistency
models, or other low-level details. Furthermore channels come in many flavors:
bounded, unbounded, blocking, non-blocking, and more. Channels can be implemented using
shared memory, with lock or lock-free queues. Channels have comparable performance to the
lock and shared memory approach.

\subsection{Intermittent Failures In Servo}
Servo \cite{servo} is a parallel browser engine with the goals: highly concurrent,
memory safe, and easily embeddable. Servo achieves these goals thanks to
its use of Rust: a novel systems programming language with stronger memory safety,
data race freedom, and static guarantees. As well as featuring a concurrent message passing
architecture. This allows Servo to be highly concurrent, maximizing parallelism,
while maintaining its components decoupled.

Nondeterminism is inherent to concurrent systems. Servo is no exception. Adding to
the challenges of programming parallel systems, debugging is more difficult than in
sequential programs. This difficulty is further exacerbated by nondeterministic
interleaving of threads.

Servo runs a large suite of browser-agnostic tests known as web-platform-tests (WPT)
as part of the continuous integration system. WPT tests are expected to pass before
changes to the code base are accepted. Unfortunately, Servo has hundreds of WPT tests that
fail randomly for unknown reasons, known as intermittent failures. Since many intermittent
tests seemingly pass or fail at random. There are unknown sources of nondeterminism in
Servo which cause intermittent failures.

\subsection{Existing Record-and-Replay Solutions}
Record-and-replay has been widely studied in academia and used in industry. One
such state-of-the-art system is Mozilla's \texttt{rr} project \cite{rr}, which serves
as a robust RR debugging framework. It is successfully used in production
with code-bases consisting of millions of lines of code like the Firefox web browser.
At a high-level, \texttt{rr} works by intercepting nondeterministic OS system calls and recording
their values, as well as recording thread and process scheduling so later it can faithfully
replay them. During replay, \texttt{rr} uses its log to replay the system calls with the same
values seen during record, while dynamically handling scheduling to schedule threads
and processes in the same order. \texttt{rr} can fully determinize any program,
so it must record a large amount of information about an execution, so its record logs
end up being quite large. Furthermore \texttt{rr} sequentializes execution of threads and
processes, incurring particularly high performance overhead on parallel workloads.

While in principal \texttt{rr} is able to solve the problem with intermittent failures in Servo,
we believe there can be a far more lightweight, high-level approach.
\texttt{rr}'s implementation is OS dependent as it relies on many low-level OS details.
Lightweight RR attempts tame some of the nondeterminism inherent to concurrent systems
while operating at a much higher level: channel communication and thread scheduling.
All while preserving parallelism in programs, and offering low overhead.

\section{Lightweight RR}
Lightweight RR records and replays multi-threaded systems
where communication and coordination happens mainly through channel communication.
Lightweight RR does not attempt to provide faithful full program record and replay. Instead,
it lowers the amount of nondeterminism inherent to concurrent systems. This trade off
allows lightweight RR several benefits: the implementation is far simpler than
a full-fledged RR system, parallelism is maintained, only a tiny fraction of program
execution is recorded making the implementation fast and the logs small.

Lightweight RR handles the following sources of nondeterminism:
\begin{itemize}
\item \textbf{Message arrival order}. While messages sent by a thread down a channel are guaranteed to arrive in FIFO order, multiple threads may be sending messages to the same channel simultaneously. Thus, message arrival order can vary.
\item \textbf{Select operation}. Channel implementations often have a \texttt{select} operation. A \texttt{select} allows for receiving one or more message(s) from any channel with a message at the time of the operation, or blocking until a message arrives. Which messages are ready on a given select is timing dependent, so this is a nondeterministic operation.
\end{itemize}

Our approach is guaranteed to successfully replay an execution given that the above are the only sources of nondeterminism. If programs have nondeterminism outside the scope of lightweight RR, e.g. network communication, IO, thread locking, those programs may not replay successfully. If the replay execution diverges from the recorded execution, we say the execution has \textit{desynchronized}.

Lightweight RR is robust to desynchronizations. When desynchonization is detected, the lightweight RR stops replaying the recorded execution and falls back to allowing the program to run natively.

\subsection{Lightweight RR for Intermittent Failures}
Lightweight RR is designed to work in the context of software tests with intermittent
failures.
Any test that nondeterministically return: unexpected results, timeout, or crash, can
benefit from lightweight RR. First, we must capture a execution of the program with an
``expected'' results, this may take many tries depending on the test.
These ``expected'' logs are stored as part of the testing infrastructure. Later the
tests run using lightweight RR in place of the standard channel library. The test run
in replay mode where lightweight RR replays the ``expected'' execution. This allows
lightweight RR to lower the nondeterminism and thus the number of intermittent failures
for programs.

\subsection{Supported Operations}
For any message-passing channel implementation, we define the \textit{receiver} as the reading end, and the \textit{sender} as the writing end. We assume the channels are unbounded (do not have a maximum capacity), and may support multiple producers but a single consumer (MPSC). These assumptions fit the needs of most programs but nothing fundamentally stops us from supporting bounded, or multiple producers multiple consumer (MPMC) channels.

Lightweight RR supports all common operations on channels, including:
\begin{itemize}
\item  \textbf{receive}: Block until we receive a message from a sender.
\item  \textbf{select}: Block on one or more receivers until one or more message(s) arrives from any channel.
\item  \textbf{try\_receive}: Non blocking variant of receive. If no message is available, an
  error is returned.
\item  \textbf{timeout\_receive}: blocks until a message arrives or the specified time elapses.
\item  \textbf{send}: non-blockingly send a message through a channel.
\end{itemize}

\subsection{What is lightweight?}
Lightweight RR does not reinvent the wheel by implementing deterministic channels from
scratch, instead it is implemented as a wrapper around an existing channel
implementation.

Furthermore, unlike fully deterministic approaches to RR, lightweight RR is
implemented at the language-level as a user library, agnostic to the operating system and
hardware details.

We highlight the following advantages between traditional RR methods and lightweight RR:
\begin{itemize}
\item Implemented as a language-level library, portable across operating systems and
  computer architectures.
\item Threads run in parallel.
\item Only channel communications events are recorded. No program data is recorded.
  Leading to smaller record logs.
\item Robust to divergence of program execution during replay.
\item The implementation wraps an existing channel implementation. Thus, the
  implementation is a straightforward transform.
\end{itemize}

\section{Design}
\subsection{Assigning Deterministic IDs}
In order to faithfully replay execution of a program, we must assign a deterministic
thread ID (DTI) to all threads. DTIs must be unique and deterministic
even in the face of racy thread spawning. Starting from the main thread,
every thread spawn event is assigned an ID equal to the number of children that
have spawned from this thread so far $child\_id$. A single counter per thread keeps tracks of this information. Then, a DTI is generated as a vector of $child\_id$s from the main thread, down to the new thread. The main thread has a DTI of \texttt{[]} (an empty list). As an example, the first thread spawned by the main thread is assigned a DTI of \texttt{[1]}, while the second great-grandchild of the main thread would have a DTI of \texttt{[1, 1, 2]}. Therefore, the length of a DTI is equal to the depth of the thread tree.

Analogously, every channel receiver/sender pair is assigned a unique, deterministic channel ID (DCI).
The DCS consists of a \texttt{(DTI, N)}, where \texttt{N} is a per-thread
counter, which is increased after generating a DTI.

DTIs are necessary for correctly replaying executions. Meanwhile DCIs are useful to detect
desynchronizations, debugging, as well as understanding complex program interactions among
threads, channels, and messages.

\subsection{Recording Events}
Given a channel message of type $T$, we transform it to a tuple \texttt{(T, DTI)}. Where \texttt{DTI} is
the DTI of the thread who sent this message. That is, a thread now sends its DIT down
the channel along with the message, this allows us to track which thread the message
originally came from.

During record mode, we let the program run with as little changes to the execution as
possible. When the program calls a channel function, we allow the function to go through, and merely record some information necessary for replay later.

For all channel communication events, we record:
\begin{itemize}
\item Event Id: A per-thread integer that increases by one on every channel communication
  event. This should be thought of as the \textit{logical time} that an event occurred at.
\item Sender ID: The DTI of the sender of a message.
\item Event Type: \texttt{send}, \texttt{receive}, \texttt{try\_receive}, \texttt{select}, etc.
\item Data Type: The type of data sent down the channel, e.g. string, integer, bool, etc.
\item Channel Flavor: The channel variant for an event, e.g. \texttt{ipc-channel}, bounded channel,
  \texttt{crossbeam channel}, etc.
\item Event Status: Possible return variant of an event (see more below).
\item DCI: The ID(s) of the channel(s) involved in this event.
\end{itemize}
Notice our record log does not need to record the value of the data sent down the channel.

Event status are the possible return variants of an event. For example, \texttt{try\_receive}
may either: timeout, get a receive error, or successfully receive a value. In case of a
success we record the DTI of the sender thread. The sender's DTI is
important for faithfully replaying executions when there are multiple writers
to the same channel.

Select operations have an ordering to their receiver set. Each receiver is given
an index to its position on the select, based on the order in which receivers were added to
the set. On select events, we record the index (or indices for multiple events) of the
receiver(s) with messages ready. Since receivers on a select are themselves MPSC channels. We
also record the DTI of the sender for all receivers which returned a message during a select.

Many of the items recorded above are not strictly necessary. Instead, the are useful for
robust logging (debugging), and detecting desynchronizations of executions. For users
interested in running lightweight RR in production environments, many of these values could
be removed at build-time.

\subsection{Replaying Events}
While record mode tries to affect the program's execution as little as possible,
replaying requires wrangling in the program to force it to execute the same as the recorded
execution. We may have to block threads and execute multiple channel receives
per single channel receive requested by the user. All these operations are not directly
observable to the user, and lightweight RR always maintains correct semantics with respect
to the underlying channel implementation. Therefore, any program that works with a channel
implementation will work equally well with lightweight RR.

Using \texttt{(Sender ID, Event ID)} we can deterministically identify an event for any thread.
For a given event, we consult the log to see what actions are expected for this event.
We compare the event type, DCI, and channel flavor to ensure we are still
synchronized with the execution. Otherwise, a desynchronization error is returned.

Next, we consult the event status. If the status was e.g. timeout, we don't bother
doing the actual operation on the channel, and instead immediately return a timeout.
Even if the event status indicates the event succeeded, we may still not
do the actual operation. Instead, no matter the actual receive events e.g.
\texttt{timeout\_receive}, \texttt{try\_receive},
we implement a \texttt{rr\_recv()} as follows.

\texttt{rr\_recv()} call the blocking \texttt{receive()} on the receiver directly.
The blocking receive will eventually return a message of type \texttt{(DTI, T)}. We
compare the received \texttt{DTI} against the expected \texttt{DTI} from the log.
If they match,
we have found the correct message to return. Otherwise, this event came from another
thread, so we add it to a per-receiver buffer of messages. We loop on \texttt{receive()}
until the correct event is found. Next time there is a call to \texttt{rr\_recv()},
the buffer is first checked for the expected event before looping on \texttt{receive()}.

On \texttt{select}, we use the expected index/indices to retrieve the correct receiver(s)
from the select set and call \texttt{rr\_recv()} on each receiver. Notice \texttt{rr\_recv}
takes care of buffering all ``wrong'' messages from other senders, so the correct message
is automatically returned.

We must be careful with our use blocking \texttt{receive()} as a desynchronization may
cause the thread to block forever. See Section \ref{sec:handle_desync}.

\subsection{Running Off the End of the Log}
When replaying an execution, it is common to fall of the end of the log. That is,
there is no \texttt{(DTI, N + 1)} event in the log. Missing log entries usually mean
that either: the program did not reach this point of the execution during record, or a desynchronization has occurred.

The former can happen for programs that race between exiting and communicating through
channels. While one could argue this a source of nondeterminism outside the scope of
lightweight RR, doing so would render lightweight RR useless for many programs.
Instead, we handle end of the log events by blocking the thread on a conditional
variable (this simulates the behavior of the thread not reaching this event in the recorded execution). See Section \ref{sec:handle_desync} for details on when this conditional variable
is notified.

\subsection{Handling Desynchronizations}
\label{sec:handle_desync}
We handle desynchronization events robustly by continuing to execute in
a best effort mode (See Section \ref{sec:future_work} for possible extensions).
Once a desynchronization is detected, initial user input determines whether the
program should error out or continue executing the program.

Programs desynchronize when there are sources of nondeterminism outside of what lightweight
RR handles. For example network IO, timers, randomness, etc. These
other sources of nondeterminism are outside the scope of this work. We expect many
programs to have other sources of nondeterminism. So lightweight RR must be robust to
desynchronization errors.

Once we detect that execution has failed. We switch from the replay method described
above, to merely doing the operation the user has asked for. This allows us to
continue the execution but with no determinism guarantees. We notify all threads
blocked on the end-of-log conditional variable, this unblocks these threads and
they continue running on desynchronization mode. Special care must be taken to
flush any values buffered by our \texttt{rr\_recv()} operation.

If a program replay desynchronizes, threads blocking on \texttt{receive()} may never never
return. To avoid deadlocks, we use \texttt{receive\_timeout()} instead of \texttt{receive()}
in the implementation of \texttt{rr\_recv()}. If the timeout time elapses, we consider this
a type of desynchronization error and run the rest of the program in desync mode.

\section{Implementation}
We implemented lightweight RR as a Rust library \texttt{rr-channel}. Our Rust implementation
wraps two popular Rust libraries
for channel communication: \texttt{ipc-channel} and \texttt{crossbeam-channel} (it would be
trivial to support the standard library channels as well). Rust does not have
support for dependency injection or inheritance, so users must swap all instances of the
channel library they are using to \texttt{rr-channel}. To make this process seamless,
\texttt{rr-channel} exposes the exact same API as \texttt{ipc-channel} and
\texttt{crossbeam-channel}. Therefore, the only change needed is editing the name of the
dependency when it is imported.

We must assign \texttt{DTI} to all threads. So a user must also use our
\texttt{rr-channel} thread spawn function to ensure the thread receives a \texttt{DTI}. A
program may not always be able to use our thread spawn function: threads may spawn as in
program dependencies outside the control of the user. \texttt{rr-channel} handles this
case by giving these threads
a \texttt{None} \texttt{DTI}. Events are still recorded and replayed for \texttt{None}
\texttt{DTI}. However, if two threads, both with \texttt{None} \texttt{DTI}s are both
writing to the same channel, we cannot distinguish these them, so the execution may not
be deterministic. We print a warning to users if this case is detected.

It is straightforward for our \texttt{rr-channel} implementation to support
any channel implementations as long as the underlying implementation supports
one single channel operation: \texttt{try\_receive()}.

\section{Evaluation}
For our evaluation, we integrate \texttt{rr-channel} into Servo. Servo is a
highly complicated program featuring a fully concurrent, message passing architecture.
Servo represents a high target for lightweight RR: as Servo uses many complicated channels, to send many messages among its multitude of threads.
Therefore, Servo is the perfect platform to stress-test lightweight RR and get
a proper sense on the limits of our approach.

Thanks to the design of \texttt{rr-channel}, lightweight RR integration was fairly simple.
Even though Servo is split into many libraries, with dozens of decoupled components. One
day's worth of work is enough to have Servo using lightweight RR channels everywhere.

\subsection{Reducing intermittent failures}

To test the effectiveness of our approach we compiled a list of WPT tests known to
fail intermittently in Servo. Using Github API we fetched all GitHub Issues for Servo
marked as \texttt{I-intermittent}. We found 417 unique intermittently failing tests.
We ran those 417 tests 100 times (baseline) to get the average number of times a
test returned the following possible statuses:
\begin{itemize}
  \item expected: The test returned an expected outcome.
  \item unexpected: The test returned an unexpected outcome.
  \item crash: Servo crashed unexpectedly while executing this test.
  \item timeout: The test timed out before returning.
\end{itemize}

Out of those 417 tests, only 43 displayed intermittent behavior. That is, they test
would change from one state above to another. 9/43 never returned
``expected'' so they were ineligibly for lightweight RR.

To test our lightweight RR solution. We looped on the remaining 34 tests until
we captured an expected execution. 2 tests were
discarded because we were unable to record an expected execution after 100
tries. Even though these tests returned expected at least once in baseline. Note
for the 2 tests in the previous sentence, as well as the 9 tests that never returned
expected we could have recorded their unexpected executions as well, if the user
is less interested in recording the expected execution, and just wants more consistent
test results.

The remaining 32 tests make up our experimental results. Using the expected execution
logs, we ran these tests 100 times in replay using our modified Servo.

\begin{figure}
  \includegraphics[width=\linewidth]{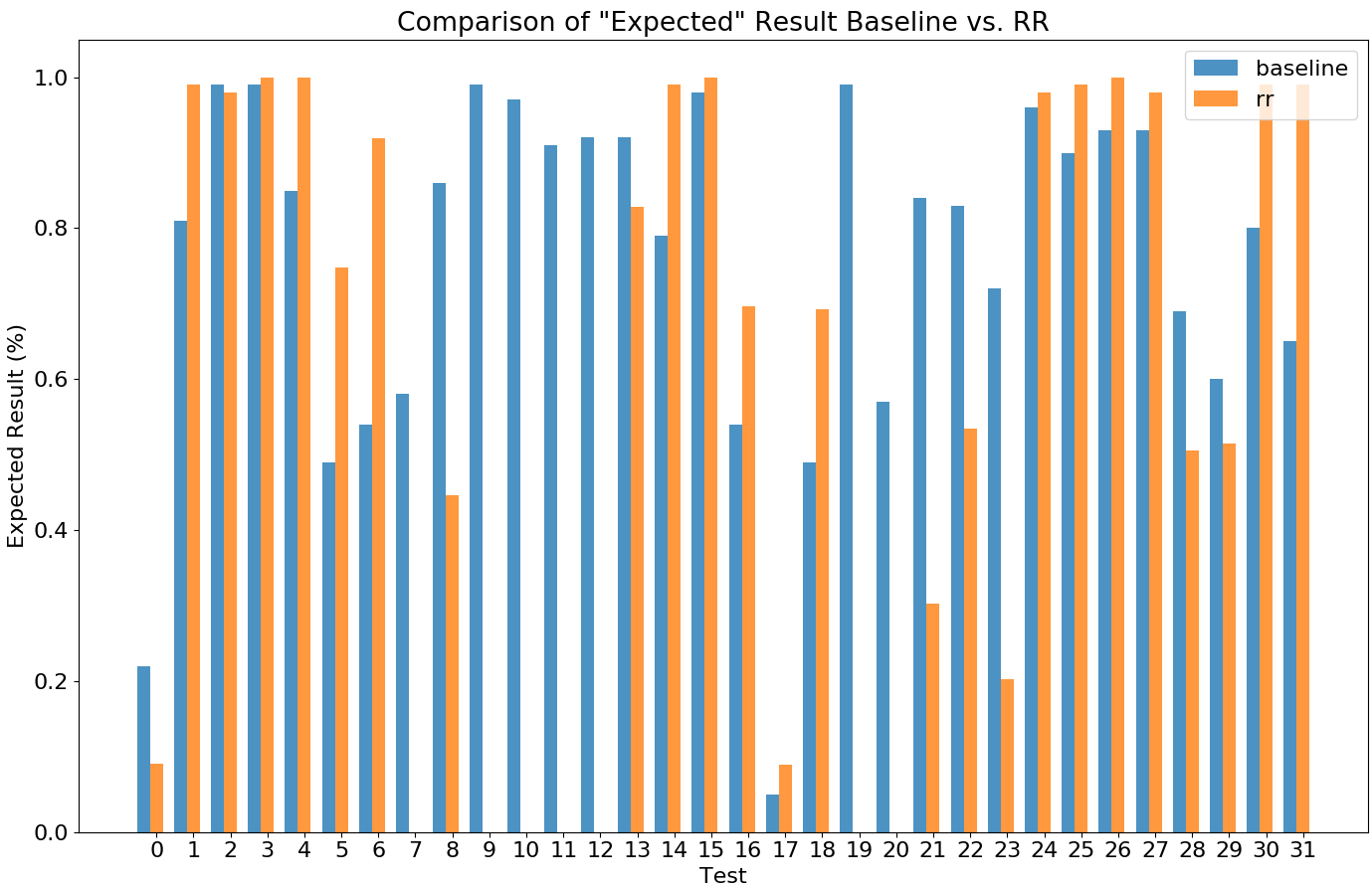}
  \caption{Results of comparing baseline with lightweight RR. Tests where rr
    is higher than the baseline represent cases where lightweight RR improved intermittent
    test's expected times. Missing orange bars represent a timeout during replay. }
  \label{fig:result1}
\end{figure}

Figure \ref{fig:result1} shows the results for our 32 tests. The results are
not great, but promising. While lightweight RR does seem to improve results,
modestly to greatly depending on the test, sometimes lightweight RR actually returns
less expected results. Many tests also failed to run all the way and timed out
every time, these represent the missing rr bars in Figure \ref{fig:result1}. See Section \ref{sec:discussion} for discussion and interpretation of this results.

\subsection{Performance Overhead}
Lightweight RR only records channel communication events, a tiny subset of information
recorded by traditional RR methods. Therefor we expect performance overhead to be quite
smaller than traditional solutions.

\begin{figure}
  \includegraphics[width=\linewidth]{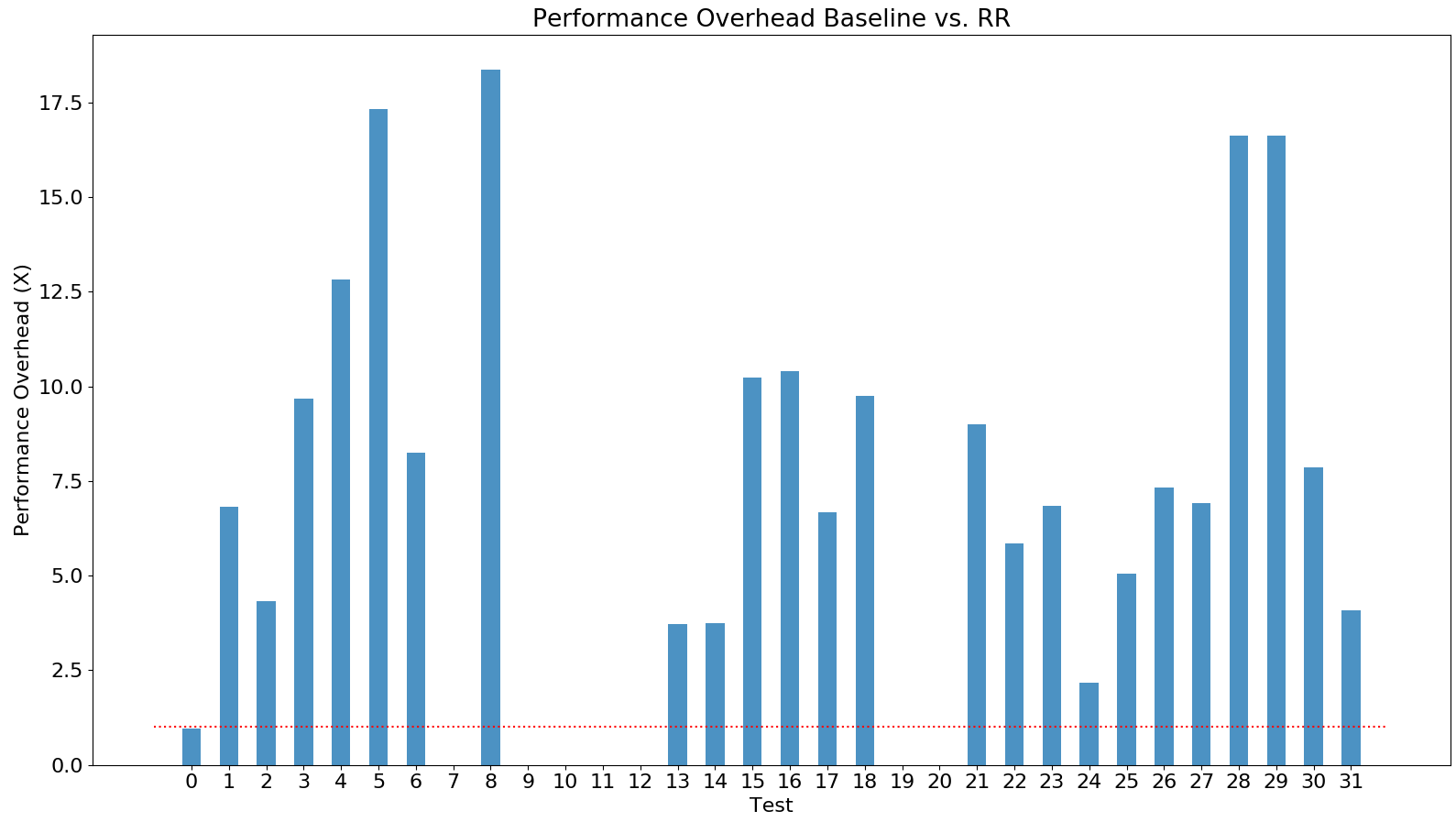}
  \caption{Performance of lightweight RR over baseline. Missing entries represent
  tests where rr-channels failed to run before timeout.}
  \label{fig:result_timing}
\end{figure}

We timed the our 32 tests using Servo's testing infrastructure reported elapsed time
per test. The average time per tests is based off 100 executions of each test. Both
baseline and lightweight RR were measured in this manner. Figure \ref{fig:result_timing}
represents these results. As can be seen in this graph, the performance is terrible.
We have identified the possible source of slowdown and hope to rerun the experiments
soon.

\subsection{Space overhead}
For the evaluation, the logs were written out using a compact, binary serializer.
Out of 30 recorded logs from our tests above, the median log size was 78.5KB.
A more in depth comparison would be needed to compare log sizes versus existing
systems like \texttt{rr}. However, we believe the log sizes are small enough to
not add a storage burden, even for test suites with tens of thousands of tests.

\section{Discussion}
\label{sec:discussion}
The current evaluation results are disappointing to say the least. Here we discuss
some of the reasons why lightweight RR does not work well in its current state.

Several tests failed to run all the way in replay mode, and instead timed out. We attribute
these failures to logic bugs in the implementation. We believe, with further implementation
effort, lightweight RR should never deadlock.

Servo is a complicated, highly concurrent program. Therefore, Servo represents a high
mark to hit for lightweight RR. Simpler applications may fare better as far as lightweight
RR is concerned.

Servo may be far too nondeterministic
for lightweight RR to be useful: Servo may be desynchronizing early on, so the rest of
the program runs without much benefits from lightweight RR.
See Section \ref{sec:future_work} for proposed solutions to this
problem. Servo has known sources of nondeterminism we are not capturing. Some examples
include \texttt{rayon}, a library for high-performance data-parallelism. Since \texttt{rayon} threads are
not spawned through our thread spawn wrapper, these threads are not assigned DTIs. Therefore any message which comes from a \texttt{rayon} thread cannot be properly determinized.
We have observed such messages coming from \texttt{rayon} threads.
With further implementation it should be possible to assign deterministic thread
identifiers to Servo threads. Similarly, Servo uses \texttt{Tokio} library for asynchronous programming. Threads spawned through \texttt{Tokio} are not determinized.
Even handling \texttt{Tokio} and \texttt{rayon} threads, Servo naturally has many nondeterministic IO operations. Network traffic, disk IO, and timers, all contribute to nondeterministic executions in Servo.

It may just be the case that Servo is too nondeterministic in its current state to benefit
from our approach. Perhaps a combination of lightweight RR and some small modifications
of Servo components may yield better results.

\section{Future Work}
\label{sec:future_work}
The current work just barely scratches the surface: providing an implementation
and doing a first attempt experiment. We believe with some iteration we could significantly
increase how well lightweight RR works.

Lightweight RR has many possible extensions and future work. Currently, when the replay
execution desynchronizes, we merely fall back to executing the program natively. A more
elaborate desynchronization recovery scheme could be robust to some desynchronized
events. Some threads could be desynchronized while other threads continue to run
synchronized. Similarly, we could look at real synchronization recovery by attempting
to match the execution back at a later point.

Once a desynchronization is detected, it is indicative of a source of nondeterminism
in some program thread. Using the current execution, and the log, it may be possible
to help the programmer narrow down and identify the source of nondeterminism.

When we detect a desynchronization, we could start recording this alternate program
execution. Later the two executions could be compared to better understand why
they diverged. Furthermore, when considering continuous integration, and as programs
evolve overtime, we expect changing part of the code will have changes in the execution,
therefore, as new patches are tested for merging we could record the new executions,
and use those for future record and replay testing.

The log we record is a full record of all communications and thread topology for a
given execution. This log could create a visualization of threads, and channel
communication for the program execution. This is extremely useful to a developer as
a visualization tool to understand concurrent programs.

Servo is careful to use non-blocking, unbounded, channels. This helps Servo developers
reason about deadlock-freedom, as long as there is no cycles in the communications among
threads
of Servo. Using the recorded logs, we could verify that cyclic dependencies don't exist
for any given execution of Servo, thus increasing confidence in deadlock-freedom.

The evaluation is currently incomplete. More work is required to understand what kinds
of applications or tests benefit most from lightweight RR (if any). Performance is also
beyond disappointing, work is needed to understand performance implications of lightweight
RR.

\section{Conclusion}
In this paper, we describe lightweight RR. A halfway point between heavyweight, fully
deterministic systems, and no determinism enforcement. Lightweight RR is designed to
empirically lower the amount of intermittent test failures in a program's test suite,
while enjoying minimal performance overhead and record-log sizes. Lightweight
RR has exciting and promising future work avenues beyond the just record and replay.
With more work, lightweight RR could prove to be a useful tool for the development
and debugging of concurrent systems.

\bibliography{bibliography}{}

\begin{thebibliography}{1}
\providecommand{\url}[1]{#1}
\csname url@samestyle\endcsname
\providecommand{\newblock}{\relax}
\providecommand{\bibinfo}[2]{#2}
\providecommand{\BIBentrySTDinterwordspacing}{\spaceskip=0pt\relax}
\providecommand{\BIBentryALTinterwordstretchfactor}{4}
\providecommand{\BIBentryALTinterwordspacing}{\spaceskip=\fontdimen2\font plus
\BIBentryALTinterwordstretchfactor\fontdimen3\font minus
  \fontdimen4\font\relax}
\providecommand{\BIBforeignlanguage}[2]{{%
\expandafter\ifx\csname l@#1\endcsname\relax
\typeout{** WARNING: IEEEtran.bst: No hyphenation pattern has been}%
\typeout{** loaded for the language `#1'. Using the pattern for}%
\typeout{** the default language instead.}%
\else
\language=\csname l@#1\endcsname
\fi
#2}}
\providecommand{\BIBdecl}{\relax}
\BIBdecl

\bibitem{servo}
\BIBentryALTinterwordspacing
``Servo, the parallel browser engine project.'' [Online]. Available:
  \url{https://servo.org/}
\BIBentrySTDinterwordspacing

\bibitem{rr}
\BIBentryALTinterwordspacing
R.~O{\textquoteright}Callahan, C.~Jones, N.~Froyd, K.~Huey, A.~Noll, and
  N.~Partush, ``Engineering record and replay for deployability,'' in
  \emph{2017 {USENIX} Annual Technical Conference ({USENIX} {ATC} 17)}.\hskip
  1em plus 0.5em minus 0.4em\relax Santa Clara, CA: {USENIX} Association, Jul.
  2017, pp. 377--389. [Online]. Available:
  \url{https://www.usenix.org/conference/atc17/technical-sessions/presentation/ocallahan}
\BIBentrySTDinterwordspacing

\end{thebibliography}
\bibliographystyle{IEEEtran}

\end{document}